\documentclass[aps,pra,twocolumn,showpacs,superscriptaddress]{revtex4}
\usepackage{graphicx}
\usepackage{amsmath}
\usepackage{amssymb}
\usepackage{lscape}

\input{epsf}
\begin{document}

\title{Three-body bound states in a harmonic waveguide with cylindrical
symmetry}

\author{D. Blume}
\affiliation{Department of Physics and Astronomy,
Washington State University,
  Pullman, Washington 99164-2814, USA}
\affiliation{ITAMP, Harvard-Smithsonian Center
for Astrophysics, 60 Garden Street, Cambridge, Massachusetts 02138, USA}

\date{\today}

\begin{abstract}
Highly-elongated quasi-one-dimensional cold atom
samples have been studied extensively over the past years
experimentally and theoretically.
This work determines the energy spectrum of two identical fermions and a 
third distinguishable particle as functions
of the mass ratio $\kappa$ and the free-space $s$-wave scattering
length $a_{3\text{D}}$ between the identical fermions and the distinguishable
third particle in a cylindrically symmetric
waveguide
whose symmetry axis is chosen to be along the $z$-axis.
We focus on the regime where the mass of the identical fermions is
equal to or larger than that of the third distinguishable particle.
Our theoretical framework accounts explicitly for the motion along
the transverse confinement direction.
In the regime where excitations in the transverse direction are
absent (i.e., for states with projection quantum number $M_{\text{rel}}=0$),
we determine the binding energies for states with odd parity in $z$.
These full three-dimensional energies deviate significantly
from those obtained within a strictly one-dimensional
framework when the $s$-wave scattering length
is of the order of or smaller than the oscillator length
in the confinement direction.
If transverse excitations are present, we predict the
existence of a new
class of universal three-body bound states with $|M_{\text{rel}}|=1$
and positive parity in $z$. These bound states arise on the positive
$s$-wave scattering length side if the mass ratio
$\kappa$ is sufficiently large.
Implications of our results for ongoing cold atom experiments are discussed.
\end{abstract}


\maketitle

\section{Introduction}
Ultracold Bose and Fermi gases provide a unique environment for
exploring few-body 
physics~\cite{doertereview,braatenReview,chrisPhysToday,rudiTrend}. 
In the ultracold regime, the de Broglie wave length is much larger
than the range of the underlying two-body potential, which implies that
the details of the two-body interactions are, to a good approximation,
negligible. To leading order, the interactions between 
two particles can be described by a single atomic
physics parameter, the free-space $s$-wave scattering length
$a_{3\text{D}}$.
For a large number of atomic species,
the $s$-wave scattering length can be tuned to essentially
any value experimentally by varying an external magnetic field
in the vicinity of a magnetic Fano-Feshbach 
resonance~\cite{chin}.
The ability to tune the $s$-wave scattering length to large positive
and negative values, or even to zero, has opened the possibility to
systematically map out the system behavior from the non-interacting regime
to the weakly-attractive (weakly-repulsive) regime to the
strongly-attractive (strongly-repulsive) 
regime~\cite{doertereview,giorginireview,blochreview}.

The fact that the de Broglie wave length is,
in the ultracold regime,  much larger than the
van der Waals length of atom-atom interactions justifies important
simplifications in the theoretical treatment of cold atom gases.
Specifically,
the true atom-atom potential, which typically supports many two-body
bound states, can be replaced by a simple model potential such as
a zero-range
pseudo-potential or a Gaussian potential, which supports at most a few
two-body bound states.
If the free-space scattering length of the model potential
agrees with that of
the true atom-atom potential, then theoretical
treatments that utilize a model potential are, in general,
expected to describe the low-energy physics with good accuracy.

This work determines the bound state spectrum of two identical
fermions with mass $m_h$ and a third distinguishable
particle with mass $m_l$
in a harmonic waveguide with cylindrical symmetry. 
The identical fermions interact through a simple two-body model potential
with $s$-wave scattering
length $a_{3\text{D}}$ with the third distinguishable particle.
Since the scattering between the identical fermions is, away from a $p$-wave
resonance or
higher partial wave resonances, suppressed by the Wigner threshold law,
we assume that the identical fermions do not interact.
We determine the bound state properties of this three-body system
as functions of the interspecies $s$-wave scattering length $a_{3\text{D}}$
and the mass ratio $\kappa$, where $\kappa=m_h/m_l$;
we consider the regime $1 \le \kappa \le 12$.
The bound state
properties of fermionic three-body systems with unequal masses
have previously been investigated in mixed dimensions~\cite{nishidamixed,peng}.
While atomic three-body systems in free space share many
characteristics with the low-energy properties of few-nucleon
systems~\cite{nuclear},
three-atom systems in a harmonic waveguide with cylindrical
symmetry have no direct nuclear analog.

If transverse excitations 
are absent (i.e., if $M_{\text{rel}}=0$) and if
the size of the three-body bound state is much larger than
the harmonic oscillator length $a_{\text{ho}}$ that characterizes the
confinement in the transverse direction,
then a strictly one-dimensional Hamiltonian with 
appropriately chosen one-dimensional coupling constant 
provides a qualitatively correct 
description~\cite{olshanii,moore,mora,mora-2005,gharashi}. 
However, when the 
size of the three-body bound state becomes comparable to or smaller than
$a_{\text{ho}}$, then the trimer ``feels'' the full three-dimensional space
and we find, in agreement with 
what might be expected naively, that a simple effective one-dimensional
Hamiltonian provides a poor description.
We also investigate the properties of states with $|M_{\text{rel}}|=1$
and positive parity in $z$.
This case has, to the best of our knowledge,
not been considered in the literature.
At first sight, it may seem that the excitation in the transverse 
direction would prevent the formation of three-body bound states.
Indeed, this is the case for mass ratios
not much larger than one. For sufficiently large $\kappa$
and positive $s$-wave scattering length,
however, the attraction is sufficiently large to ``outweigh''
the energy increase due to the projection
quantum number $M_{\text{rel}}$ being finite.
The existence of three-body bound states
on the positive $s$-wave scattering length side is 
related to the fact that the three-body system in free
space, i.e., in the absence of the waveguide,
supports universal bound states with finite angular momentum
 if $\kappa \gtrsim 8.173$ and $a_{3\text{D}}>0$~\cite{kart07,endo11}.
Analogous effects have previously been studied in quasi-two-dimensional 
systems~\cite{levinsen,pricoupenko}.
Experimentally, the three-body bound states can potentially be probed via
radio-frequency spectroscopy or detected via loss features due
to three-body recombination processes. For K-Li 
mixtures~\cite{coldatoms1,coldatoms2,coldatoms3,coldatoms4,coldatoms5}, 
e.g., the
three-body bound states should have profound effects on the system dynamics.

The remainder of this paper is organized as follows.
Section~\ref{sec_theory} outlines the theoretical framework.
Specifically, Sec.~\ref{sec_hamiltonian} introduces the system
Hamiltonian and discusses its symmetry properties;
Sec.~\ref{sec_ecg} summarizes the numerical approach used
to obtain the three-body spectra;
and
Sec.~\ref{sec_twobodyreview} reviews a number of key results for 
two particles in a waveguide geometry.
Section~\ref{sec_results}
discusses our results for different symmetries.
Energy spectra are presented
and the dependence of the energies on the range
of the underlying two-body potential is analyzed.
Lastly, Sec.~\ref{sec_summary}
summarizes.

\section{Theoretical framework}
\label{sec_theory}

\subsection{System Hamiltonian and symmetries}
\label{sec_hamiltonian}
We consider three particles with masses $m_j$ and position vectors
$\vec{r}_j=(x_j,y_j,z_j)$ in a cylindrically symmetric
waveguide with angular trapping frequency $\omega$.
Assuming isotropic interactions $V_{\text{G}}(r_{jk})$ 
($r_{jk}=|\vec{r}_j-\vec{r}_k|$)
between the distinguishable 
particles, the system Hamiltonian $H_{\text{tot}}$
reads
\begin{eqnarray}
\label{eq_hamtot}
H_{\text{tot}} = \sum_{j=1}^3 
\left(
\frac{-\hbar^2}{2m_j} \nabla_{\vec{r}_j}^2 + \frac{1}{2} m_j \omega^2
\rho_j^2 \right) 
+ \sum_{j=2}^3 V_{\text{G}}({r}_{1j}),
\end{eqnarray}
where $\rho_j^2 = x_j^2+y_j^2$.
In Eq.~(\ref{eq_hamtot}), $\nabla_{\vec{r}_j}^2$ denotes the three-dimensional
Laplacian of the
$j$th particle.
Our Hamiltonian assumes that the three particles 
with masses $m_1=m_l$ and $m_2=m_3=m_h$
all feel the same angular trapping
frequency.
While this is fullfilled ``automatically'' for equal-mass systems,
for unequal-mass systems the realization of equal trapping frequencies
requires some fine-tuning~\cite{orso}.
In Eq.~(\ref{eq_hamtot}), 
$V_{\text{G}}$
denotes a Gaussian model interaction potential with range $r_0$ and 
depth $V_0$ ($V_0>0$),
\begin{eqnarray}
V_{\text{G}}(r)=-V_0 \exp \left[ -\left(\frac{r}{\sqrt{2}r_0} \right)^2 \right].
\end{eqnarray}
For a fixed range $r_0$, $V_0$ is adjusted such that
$V_{\text{G}}$ supports no free-space bound state for $a_{3\text{D}}<0$ and one free-space
bound state for $a_{3\text{D}}>0$. We work in the regime where
$r_0$ is much smaller than the harmonic oscillator length $a_{\text{ho}}$,
\begin{eqnarray}
a_{\text{ho}} = \sqrt{\frac{\hbar}{2 \mu \omega}},
\end{eqnarray}
where the two-body reduced mass  $\mu$ is given by $m_h m_l/(m_h+m_l)$.

To analyze the symmetry properties of $H_{\text{tot}}$,
we introduce cylindrical coordinates,
$(x_j,y_j,z_j)=(\rho_j \cos \varphi_j, \rho_j \sin \varphi_j,z_j)$.
In these coordinates, we have 
$r_j^2 = \rho_j^2 +z_j^2$ and $r_{jk}^2=\rho_{jk}^2+z_{jk}^2$,
where $\rho_{jk}^2  = (x_j-x_k)^2+(y_j-y_k)^2$ and $z_{jk}=z_j-z_k$.
It can be checked readily that 
$H_{\text{tot}}$ is invariant under a rotation about the $z$-axis and
when changing all $x_j$ coordinates to $-x_j$ (and similarly for
$y_j$ and $z_j$).
Correspondingly, we can find simultaneous eigenstates of
$H_{\text{tot}}$, the $z$-component of the
orbital angular momentum operator $L_{\text{tot},z}$,
the parity operator $P_{z}$ ($P_{z}$ sends all $z_j$ to $-z_j$), 
and the parity operator $P_{\vec{\rho}}$ ($P_{\vec{\rho}}$ 
sends all $x_j$ to $-x_j$ and all $y_j$ to $-y_j$).

Another important property of $H_{\text{tot}}$ is that it can be written as a 
sum of the relative Hamiltonian $H_{\text{rel}}$ and the center of mass
Hamiltonian $H_{\text{cm}}$,
\begin{eqnarray}
H_{\text{tot}}=H_{\text{rel}}+H_{\text{cm}}.
\end{eqnarray}
To write out $H_{\text{rel}}$ and $H_{\text{cm}}$, it is convenient to
transform to 
Jacobi coordinates
$\vec{r}_{\text{J}1}$, $\vec{r}_{\text{J}2}$ and $\vec{r}_{\text{J}3}$ 
[$\vec{r}_{\text{J}j}=(x_{\text{J}j},y_{\text{J}j},z_{\text{J}j})$],
where 
\begin{eqnarray}
\vec{r}_{\text{J}1} = 
\vec{r}_1 - \vec{r}_2 
,
\end{eqnarray}
\begin{eqnarray}
\vec{r}_{\text{J}2} = 
\frac{m_1 \vec{r}_1 + m_2 \vec{r}_2}{m_1+m_2} - \vec{r}_3
\end{eqnarray}
and 
\begin{eqnarray}
\vec{r}_{\text{J}3} = 
\frac{m_1 \vec{r}_1 + m_2 \vec{r}_2 + m_3 \vec{r}_3}
{m_1+m_2+m_3}
.
\end{eqnarray}
The center of mass Hamiltonian can be written in
terms of $\vec{r}_{\text{J}3}$ 
and the 
relative Hamiltonian $H_{\text{rel}}$ 
in terms of
$\vec{r}_{\text{J}1}$ and $\vec{r}_{\text{J}2}$.

In the following, we focus on solving the relative Schr\"odinger equation
\begin{eqnarray}
\label{eq_releq}
H_{\text{rel}} \Psi(\vec{r}_{\text{J}1},\vec{r}_{\text{J}2}) 
= E_3 \Psi(\vec{r}_{\text{J}1},\vec{r}_{\text{J}2})
\end{eqnarray}
for the eigenstates $\Psi$ with eigenenergy $E_3$.
As before, we employ cylindrical coordinates, i.e.,
we write
$\vec{r}_{\text{J}j} = 
(\rho_{\text{J}j} \cos \varphi_{\text{J}j},\rho_{\text{J}j} \sin \varphi_{\text{J}j},z_{\text{J}j})$
($j=1$ and $2$).
To take advantage of the Hamiltonian's symmetry, we perform an
additional coordinate transformation, namely, we replace $\varphi_{\text{J}1}$
and $\varphi_{\text{J}2}$ by
$\Phi$ and $\phi$,
\begin{eqnarray}
\label{eq_angle1}
\Phi = \frac{1}{2} \left(
\varphi_{\text{J}1}+ \varphi_{\text{J}2} \right)
\end{eqnarray}
and
\begin{eqnarray}
\label{eq_angle2}
\phi = 
\varphi_{\text{J}1}- \varphi_{\text{J}2}.
\end{eqnarray}
It can be checked readily that
the interaction potential is independent
of the angle $\Phi$. It follows that the relative 
wave function factorizes,
\begin{eqnarray}
\Psi(\rho_{\text{J}1},\rho_{\text{J}2},\phi,\Phi,z_{\text{J}1},z_{\text{J}2})= \nonumber \\
\psi_{M_{\text{rel}}}(\rho_{\text{J}1},\rho_{\text{J}2},\phi,z_{\text{J}1},z_{\text{J}2}) \exp(i M_{\text{rel}} \Phi),
\end{eqnarray}
where $M_{\text{rel}}=\cdots,-2,-1,0,1,2,\cdots$.
For $M_{\text{rel}} \ne 0$, each eigenenergy is twofold degenerate due to the 
$M_{\text{rel}}$
quantum number.

In the following, we label our solutions by the quantum numbers
$\Pi_{\vec{\rho}}$, 
$M_{\text{rel}}$ and $\Pi_z$, 
which are defined through the action of the operators
$P_{\vec{\rho}}$, 
$L_{\text{rel},z}$, and $P_z$
on the eigenfunctions $\Psi$,
\begin{eqnarray}
P_{\vec{\rho}} \Psi = \Pi_{\vec{\rho}} \Psi,
\end{eqnarray} 
\begin{eqnarray}
L_{\text{rel},z} \Psi = \hbar M_{\text{rel}} \Psi,
\end{eqnarray} 
and
\begin{eqnarray}
P_{z} \Psi = \Pi_{z} \Psi.
\end{eqnarray} 
One finds $\Pi_{\vec{\rho}}=\pm 1$ 
[in fact, $\Pi_{\vec{\rho}}=(-1)^{M_{\text{rel}}}$]
and $\Pi_{z}=\pm 1$.
For $M_{\text{rel}}=0$, we can find simultaneous eigenfunctions
of $H_{\text{rel}}$, $P_{\vec{\rho}}$, $L_{\text{rel},z}$, $P_z$
and $A_y$, 
where the reflection operator $A_y$ sends all $y_j$ to 
$-y_j$~\cite{footnotereflection}.
Specifically, for $M_{\text{rel}}=0$, we have
\begin{eqnarray}
A_{y} \Psi = a_{y} \Psi
\end{eqnarray} 
with
$a_{y}=\pm 1$.
We determine the relative eigenenergies $E_3$
and eigenstates $\Psi$ [see Eq.~(\ref{eq_releq})] by expanding $\Psi$ 
in terms of explicitly correlated Gaussian basis functions
with good $\Pi_{\vec{\rho}}$, $M_{\text{rel}}$ and $\Pi_z$ 
(and, if $M_{\text{rel}}=0$, $a_y$)
quantum numbers
and solve a generalized eigenvalue equation.

\subsection{Explicitly correlated Gaussian basis set
expansion approach}
\label{sec_ecg}
Explicitly correlated Gaussian basis functions
have been shown to provide
accurate descriptions
of strongly-correlated systems such as
nuclei, molecules,
atoms, and quantum dots~\cite{cgbook,RMPreview}.
They have also been employed to characterize small
dilute atomic 
gases~\cite{RMPreview,sore05,stec07c,debraj}.
We write
\begin{eqnarray}
\label{eq_cgexpansion}
\Psi(\vec{r}_{\text{J}1},\vec{r}_{\text{J}2})=
{\cal{A}}  \sum_{k=1}^{N_b} c_k 
f_k(x_{\text{J}1},y_{\text{J}1},x_{\text{J}2},y_{\text{J}2},\underline{A}_{\rho,k},\vec{u}_{\rho,k}) 
\times \nonumber
\\
g_k(z_{\text{J}1},z_{\text{J}2},\underline{A}_{z,k},\vec{u}_{z,k})
,
\end{eqnarray} 
where ${\cal{A}}$ denotes the 
operator that ensures that the wave function
is anti-symmetric under the exchange of the two
identical fermions,
${\cal{A}}=1-P_{23}$ ($P_{23}$ exchanges particles 2 and 3).
The $c_k$ denote expansion coefficients. These
linear variational parameters are determined by solving the generalized 
eigenvalue
problem defined by the Hamiltonian and overlap matrices.
In Eq.~(\ref{eq_cgexpansion}), $N_b$ denotes the size of the 
basis set.
The functions $f_k$ and $g_k$ depend on a set of
non-linear variational parameters
through
$\underline{A}_{\rho,k}$, $\vec{u}_{\rho,k}$, $\underline{A}_{z,k}$
and $\vec{u}_{z,k}$~\cite{cgbook,svm}.
Here and in what follows,
underlined symbols denote matrices.

We consider two different functional forms for 
$g_k$, one that is characterized by
$\Pi_z=+1$ (referred to as $g_k^{(e)}$) and one that is characterized by
$\Pi_z=-1$ (referred to as $g_k^{(o)}$),
\begin{eqnarray}
g_k^{(e)}(z_{\text{J}1},z_{\text{J}2},\underline{A}_{z,k}) = 
\exp \left( -\frac{1}{2} \vec{z}_{\text{J}}^T \underline{A}_{z,k}
\vec{z}_{\text{J}} \right)
\end{eqnarray}
and 
\begin{eqnarray}
g_k^{(o)}(z_{\text{J}1},z_{\text{J}2},\underline{A}_{z,k},\vec{u}_{z,k}) = 
{v}_{z,k} \exp \left( -\frac{1}{2} \vec{z}_{\text{J}}^T 
\underline{A}_{z,k}
\vec{z}_{\text{J}} \right),
\end{eqnarray}
where $\vec{z}_{\text{J}}=(z_{\text{J}1},z_{\text{J}2})$ and $\underline{A}_{z,k}$ denotes a symmetric
$2 \times 2$ matrix.
The quantity ${v}_{z,k}$
is defined through
\begin{eqnarray}
v_{z,k} = \vec{u}_{z,k}^T \vec{z}_{\text{J}},
\end{eqnarray}
where $\vec{u}_{z,k}$ denotes a two-component vector.
The elements of the vector $\vec{u}_{z,k}$ and the elements of the symmetric
matrix $\underline{A}_{z,k}$ are treated as non-linear variational
parameters~\cite{cgbook,svm}.

The functions $f_k$ are characterized by the $M_{\text{rel}}$
quantum number.
For $M_{\text{rel}}>0$, we use~\cite{cgbook}
\begin{eqnarray}
\label{eq_frhom}
f_{k}^{(M_{\text{rel}})}
(x_{\text{J}1},y_{\text{J}1},x_{\text{J}2},y_{\text{J}2},\underline{A}_{\rho,k},\vec{u}_{\rho,k}) = 
\nonumber \\
(v_{x,k} + i v_{y,k})^{M_{\text{rel}}}
\exp \left( -\frac{1}{2} \vec{\rho}_{\text{J}}^T 
\underline{A}_{\rho,k} \vec{\rho}_{\text{J}}   \right),
\end{eqnarray}
where
$\vec{\rho}_{\text{J}}$ is a two-component vector, 
$\vec{\rho}_{\text{J}}=(\vec{\rho}_{\text{J}1},\vec{\rho}_{\text{J}2})$,
with the components being vector quantities themselves
[$\vec{\rho}_{\text{J}j}=(x_{\text{J}j},y_{\text{J}j})$].
The $2 \times 2$ matrix $\underline{A}_{\rho,k}$ is symmetric; the independent
elements of $\underline{A}_{\rho,k}$ are treated as non-linear variational 
parameters.
The quantities $v_{x,k}$ and $v_{y,k}$ are defined through
\begin{eqnarray}
v_{x,k} = \vec{u}_{\rho,k}^T \left( 
\begin{array}{c}
x_{\text{J}1} \\
x_{\text{J}2}
\end{array}
\right)
\end{eqnarray}
and
\begin{eqnarray}
v_{y,k} = \vec{u}_{\rho,k}^T \left(
\begin{array}{c}
y_{\text{J}1} \\
y_{\text{J}2}
\end{array}
\right),
\end{eqnarray}
where $\vec{u}_{\rho,k}$ is a two-component vector whose components
are treated as variational parameters.
To understand the form of the prefactor of 
$f_{k}^{(M_{\text{rel}})}$, we recall---using $x=\rho \cos \varphi$
and $y=\rho \sin \varphi$---that
\begin{eqnarray}
\left[ \rho \exp( i  \varphi ) \right]^{M_{\text{rel}}} 
= (x+iy)^{M_{\text{rel}}}.
\end{eqnarray}
The functions given in Eq.~(\ref{eq_frhom}) describe states
with $M_{\text{rel}}>0$. For the interaction model
considered in this paper, the energy depends on $|M_{\text{rel}}|$
and not the sign of $M_{\text{rel}}$.
Correspondingly, we only treat states with positive $M_{\text{rel}}$. 

To describe states with
$M_{\text{rel}}=0$ and $a_y=+1$, we use~\cite{cgbook}
\begin{eqnarray}
f_{k}^{(0,+1)}(x_{\text{J}1},y_{\text{J}1},x_{\text{J}2},y_{\text{J}2},\underline{A}_{\rho,k}) =
\exp \left( -\frac{1}{2} \vec{\rho}_{\text{J}}^T 
\underline{A}_{\rho,k} \vec{\rho}_{\text{J}}   \right),
\end{eqnarray}
where the superscript indicates the $M_{\text{rel}}$ and $a_y$ quantum numbers.
In this work, we do not report results for states with $(M_{\text{rel}},a_y)=(0,-1)$.
States with this symmetry do not support three-body bound states 
if the range $r_0$ of the two-body potential is much smaller
than $a_{\text{ho}}$.

Compact analytical expressions for
the Hamiltonian and overlap matrix elements
can be obtained using the results of Ref.~\cite{cgbook}.
Our optimization procedure of the non-linear variational
parameters is based on a semi-stochastic approach~\cite{svm}
and follows the scheme discussed in Ref.~\cite{debraj}.
Our calculations reported
in Sec.~\ref{sec_results} use between
600 and 1300 basis functions.
Each basis function is selected from around 4000-8000 trial functions.
The resulting energies provide variational upper bounds for the
exact ground state and excited state energies.
The basis set extrapolation error depends on the system parameters
and
is at the sub-percent level
or smaller.

Section~\ref{sec_results} reports energies for the regime where
$r_0$ is much smaller than $a_{\text{ho}}$. In selected cases,
the dependence of the energy on $r_0$ is investigated explicitly
and the $r_0/a_{\text{ho}} \rightarrow 0$ limit taken.
From a numerical point of view, the presence of three,
often vastly different, length scales
(i.e., the harmonic oscillator length $a_{\text{ho}}$, the range $r_0$
of the two-body potential, and the size of the bound state in the
$z$-direction)
is, in general, challenging.
It has been shown in the literature that the basis functions employed in this
work provide  a reliable and efficient means to describe
cold atom systems that are characterized by different
length scales.
Alternatively, one might employ
zero-range interactions and solve the Lippmann-Schwinger 
equation~\cite{mora,mora-2005,gharashi}.

\subsection{Review: Two particles in a harmonic
waveguide with cylindrical symmetry}
\label{sec_twobodyreview}
To place the three-body study into context, this section reviews a number 
of key results for the two-body system in a harmonic
waveguide with cylindrical symmetry.
The system Hamiltonian is given by Eq.~(\ref{eq_hamtot}) with
the first and second sum in Eq.~(\ref{eq_hamtot})
running from $j=1$  to $2$ and from $j=2$ to $2$, respectively  (instead of
from $j=1$ to 3 and from $j=2$ to 3).
Separating off the center of mass motion, the problem reduces
to that of a reduced mass particle that feels
a spherically symmetric 
short-range potential with $s$-wave scattering length $a_{3\text{D}}$.
In the zero-range limit, i.e., for $r_0=0$,
the two-body scattering and bound state solutions have been determined
analytically in the seminal work by Olshanii~\cite{olshanii}.
Importantly, if the one-dimensional coupling constant $g$ (see below)
and the relative two-body energy are, for a fixed
interaction potential, scaled by
$\hbar \omega a_{\text{ho}}$ and $\hbar^2/(2 \mu a_{\text{ho}}^2)$, respectively,
then these quantities are independent of the mass ratio
$\kappa$.

\begin{figure}
\vspace*{+.5cm}
\includegraphics[angle=0,width=65mm]{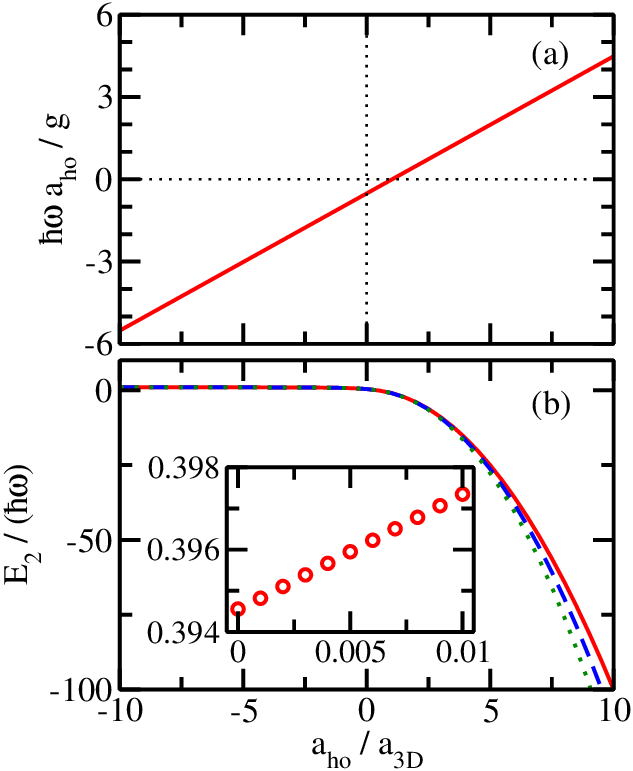}
\vspace*{0.5cm}
\caption{(Color online)
Coupling constant and binding energy for two particles
in a harmonic waveguide with cylindrical symmetry.
(a) The solid line shows the inverse 
of the
effective one-dimensional coupling constant $g$
[see Eq.~(\ref{eq_g1d})] 
as a function of
$a_{\text{ho}}/a_{3\text{D}}$ for $r_0=0$.
(b) The solid line shows
the relative two-body binding energy 
$E_{2}$ 
[see Eq.~(\ref{eq_edimer})] as a function of
$a_{\text{ho}}/a_{3\text{D}}$ for $r_0=0$.
For $a_{\text{ho}}/a_{3\text{D}} \rightarrow -\infty$,
$E_{2}$ approaches $\hbar \omega$.
For comparison, dashed and dotted lines show $E_{2}$ for
 $r_0=0.005a_{\text{ho}}$ and $0.01a_{\text{ho}}$, respectively.
Inset: Symbols show $E_{2}/(\hbar \omega)$ as a function of $r_0/a_{\text{ho}}$
for $a_{\text{ho}}/a_{3\text{D}}=0$.
}
\label{fig1}
\end{figure}

The outcome of a scattering event 
between the two
particles 
in the $(\Pi_{\vec{\rho}},M_{\text{rel}},\Pi_z,a_y)=(+1,0,+1,+1)$ 
channel 
is, for $r_0=0$, characterized by the
effective one-dimensional even parity coupling constant 
$g$~\cite{olshanii},
\begin{eqnarray}
\label{eq_g1d}
\frac{g}{\hbar \omega_{\rho} \; a_{\text{ho}}} =
 \frac{2a_{3\text{D}}}{a_{\text{ho}}} \left( 1+ \frac{\zeta(1/2)}{\sqrt{2}} 
\frac{a_{3\text{D}}}{a_{\text{ho}}} \right)^{-1}, 
\end{eqnarray}
where $\zeta(1/2)\approx -1.46035$.
Equation~(\ref{eq_g1d}) shows that the one-dimensional coupling constant 
depends on the ratio $a_{3\text{D}}/a_{\text{ho}}$. This implies that it can be tuned 
either by varying the harmonic oscillator length of the waveguide
or by varying the three-dimensional $s$-wave scattering length through
application of an external magnetic field in the vicinity of a
Fano-Feshbach resonance.
Specifically, the one-dimensional coupling constant $g$ diverges when
the $s$-wave scattering length takes the value
$a_{3\text{D}} \approx 1.03263 a_{\text{ho}}$.
The solid line in Fig.~\ref{fig1}(a) shows
the quantity $(\hbar \omega a_{\text{ho}})/g$
as a function of $a_{\text{ho}}/a_{3\text{D}}$.

While the two-body system in free space supports a weakly-bound 
state only for positive $s$-wave scattering length,
the waveguide supports a two-body bound state with 
$(\Pi_{\vec{\rho}},M_{\text{rel}},\Pi_z,a_y)=(+1,0,+1,+1)$
symmetry for all $a_{3\text{D}}$~\cite{olshanii,moore}. 
Note that the two-body system is bound if its 
relative energy
is less than $\hbar \omega$, i.e., if its relative energy is less
than the zero-point energy of the reduced mass particle
in a two-dimensional harmonic oscillator.
The relative binding energy $E_{2}$ is, for zero-range interactions,
determined by the implicit eigenvalue
equation~\cite{olshanii,moore}
\begin{eqnarray}
\label{eq_edimer}
\frac{1}{\sqrt{2}} \zeta \left( 
\frac{1}{2},- \frac{E_{2}}{2 \hbar \omega} +\frac{1}{2}
\right)
= - \frac{a_{\text{ho}}}{a_{3\text{D}}},
\end{eqnarray}
where $\zeta(\cdot,\cdot)$ denotes the Hurwitz zeta function.
The solution is shown by the solid line in Fig.~\ref{fig1}(b) as a function
of $a_{\text{ho}}/a_{3\text{D}}$.
For comparison, dashed and dotted lines show the 
relative binding energy 
for the Gaussian model potential with 
$r_0=0.005a_{\text{ho}}$ and $0.01a_{\text{ho}}$, respectively.
These two-body binding energies are obtained by solving the 
relative two-dimensional
Schr\"odinger equation using B-splines and 
are used in Sec.~\ref{sec_results}
to analyze the three-body spectra.
The finite-range effects increase as $a_{\text{ho}}/a_{3\text{D}}$ increases.
For $a_{\text{ho}}/a_{3\text{D}}=10$, e.g., the finite-range energies
deviate from the zero-range energy by $-11$\% and $-24$\%
for $r_0=0.005a_{\text{ho}}$ and $r_0=0.01a_{\text{ho}}$, respectively.
The inset of Fig.~\ref{fig1}(b) shows 
the two-body binding energy as a function of $r_0/a_{\text{ho}}$
for $a_{\text{ho}}/a_{3\text{D}}=0$. For this scattering length,
the finite-range energies lie (slightly) above the zero-range energy.
For sufficienty small $r_0$, channels with $M_{\text{rel}}>0$ and/or $\Pi_z=-1$
do not support a two-body bound state for any $a_{3\text{D}}$.

\section{Three-body bound states}
\label{sec_results}
\subsection{General considerations}
This section summarizes our search for three-body bound states
in a waveguide with cylindrical symmetry.
Throughout, we focus on parameter combinations for which
$|a_{3\text{D}}| \gg r_0$ and $a_{\text{ho}} \gg r_0$.
As discussed in Sec.~\ref{sec_twobodyreview},
the two-body system with short-range interactions supports a bound state 
with relative energy
$E_{2}$ for all $a_{3\text{D}}$. This work investigates under which conditions 
the three-body system supports states 
that are stable with respect to
the lowest dimer plus atom threshold, i.e., whose relative energy 
$E_3$ is smaller than
$E_{2}+ \hbar \omega$.
The addition of the third particle has two effects:
{\em{(i)}}
The interaction potential $V_{\text{G}}(r_{13})$ introduces an additional
attraction.
{\em{(ii)}} 
The fact that the three-particle wave function has to be
anti-symmetric under the exchange of particles 2 and 3 introduces
an effective repulsion.
Whether or not three-body bound states exist is determined by
the interplay of these two effects.

For equal masses, the three-body bound states 
in a waveguide have been characterized
in Refs.~\cite{mora,mora-2005}.
For unequal masses,
the bound states of three particles in a waveguide have 
been investigated
within a strictly one-dimensional framework, in which
the unlike particles interact through a one-dimensional
$\delta$-function potential with effective coupling 
constant $g$~\cite{dodd72,kart09,mehta14}, and not yet within a full 
three-dimensional framework.
It was found that three-body bound states are only supported if $g$
is negative, corresponding 
to $a_{\text{ho}}/a_{3\text{D}} \lesssim 1/1.03263 \approx 0.96840$,
$\Pi_z=-1$ and
$\kappa>1$. Specifically, if $\kappa$ is infinitesimally larger than
1, an infinitesimally weakly-bound three-body state emerges
for $g \rightarrow -\infty$. 
For 
$\kappa \approx 7.3791$, a second three-body bound state
becomes bound.

In addition to this one-dimensional limit, the bound states
of the three-body system in free space [i.e., in the case
where the waveguide is absent ($\omega \rightarrow 0$)] are 
known~\cite{kart07}. 
In this case,
a universal three-body bound state with $(L,{\Pi})=(1,{-1})$ 
symmetry
exists if $\kappa \gtrsim 8.173$ and $a_{3\text{D}}>0$~\cite{kart07};
here, $L$ denotes
the relative orbital angular momentum of the three-body system
and $\Pi$ the parity. In the limit that the
three-dimensional scattering length $a_{3\text{D}}$ is smaller than the harmonic oscillator
length $a_{\text{ho}}$ that characterizes the waveguide, we expect that the 
three-body solutions for the waveguide system show similarities with 
those for the
free-space system. In this limit, the waveguide can be thought of
as introducing a small
perturbation to the free-space solution. Correspondingly, we 
expect that three-body bound states exist for sufficiently large
$\kappa$, if the symmetry of the waveguide solution is ``consistent''
with the $(L,{\Pi})=(1,{-1})$ 
symmetry of the free-space solution that supports a 
universal three-body bound state.

Section~\ref{sec_kappa1} summarizes our results for $\kappa=1$.
Our full three-dimensional calculations for $\kappa=1$ confirm,
as suggested by Refs.~\cite{mora,mora-2005,kart09},
the absence of
three-body bound states 
in the $(\Pi_{\vec{\rho}},M_{\text{rel}},\Pi_z,a_y)=(+1,0,\pm 1,+1)$
channels.
Our calculations for $\kappa>1$ are summarized
in Secs.~\ref{sec_symm1}-\ref{sec_symm3}:
Section~\ref{sec_symm1} discusses
our results for the $(\Pi_{\vec{\rho}},M_{\text{rel}},\Pi_z,a_y)=(+1,0,+1,+1)$  channel,
Sec.~\ref{sec_symm2} those for the
$(\Pi_{\vec{\rho}},M_{\text{rel}},\Pi_z,a_y)=(+1,0,-1,+1)$ channel, and
Sec.~\ref{sec_symm3} those for
the
$(\Pi_{\vec{\rho}},|M_{\text{rel}}|,\Pi_z)=(-1,1,+1)$ 
channel.
We find that the 
$(\Pi_{\vec{\rho}},M_{\text{rel}},\Pi_z,a_y)=(+1,0,-1,+1)$ 
and 
$(\Pi_{\vec{\rho}},|M_{\text{rel}}|,\Pi_z)=(-1,1,+1)$ channels support three-body bound states
in certain regions of the parameter space.
These channels have an overall negative parity, i.e., 
$\Pi_{\vec{\rho}} \times \Pi_{z}=-1$, and are thus ``consistent'' with the free-space
solution that has $(L,{\Pi})=(1,{-1})$ symmetry and supports universal 
three-body bound states
for sufficiently large $\kappa$ and positive $a_{3\text{D}}$.

\subsection{Absence of three-body bound states for $\kappa=1$}
\label{sec_kappa1}
For equal masses (i.e., for $\kappa=1$), 
we find that the Pauli exclusion principle
outweighs the energy decrease due to the attraction.
Specifically, for $\kappa=1$, $r_0=0.01a_{\text{ho}}$ and 
$a_{\text{ho}}/a_{3\text{D}} \in [-10,10]$, we determined the three-body
energies using the approach discussed in Sec.~\ref{sec_ecg} and
found that
the 
$(\Pi_{\vec{\rho}},M_{\text{rel}},\Pi_z,a_y)=(+1,0, \pm 1,+1)$
and
$(\Pi_{\vec{\rho}},|M_{\text{rel}}|,\Pi_z)=(-1,1,+1)$  channels
do not support three-body bound states.
For a subset of scattering lengths, we decreased the range $r_0$ of the 
Gaussian model potential and found no significant change. 
We thus believe that three-body bound states
are absent also in the zero-range limit.

\subsection{$(\Pi_{\vec{\rho}},M_{\text{rel}},\Pi_z,a_y)=(+1,0,+1,+1)$}
\label{sec_symm1}
Considering mass ratios up to $\kappa=12$
and inverse scattering lengths
$a_{\text{ho}}/a_{3\text{D}}$ ranging from $-10$ to $10$,
we found no three-body bound states in the 
$(\Pi_{\vec{\rho}},M_{\text{rel}},\Pi_z,a_y)=(+1,0,+1,+1)$ channel.
Although the calculations were performed for a finite
range (namely for $r_0=0.01 a_{\text{ho}}$), we believe
that the results also
hold for three-body systems with zero-range interactions.
This finding is consistent with the results obtained 
within the strictly-one-dimensional
framework~\cite{kart09}. 
Moreover, the absence of 
three-body bound states 
in the $(\Pi_{\vec{\rho}},M_{\text{rel}},\Pi_z,a_y)=(+1,0,+1,+1)$ channel
for large $a_{\text{ho}}/a_{3\text{D}}$
is consistent with the fact that the free-space system 
with positive parity does not support three-body bound 
states~\cite{VNEfimov1970,EfimovLetter1972,Efimov1973,petrov}.

\subsection{$(\Pi_{\vec{\rho}},M_{\text{rel}},\Pi_z,a_y)=(+1,0,-1,+1)$}
\label{sec_symm2}
\begin{figure}
\vspace*{+.5cm}
\includegraphics[angle=0,width=65mm]{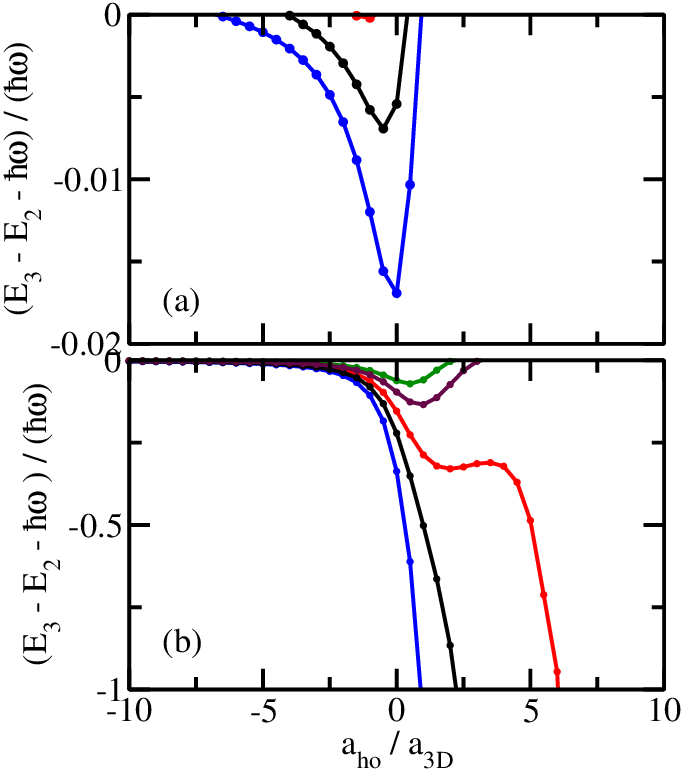}
\vspace*{0.5cm}
\caption{(Color online)
Relative three-body energies for the lowest state with
$(\Pi_{\vec{\rho}},M_{\text{rel}},\Pi_z,a_y)=(+1,0,-1,+1)$
symmetry as a function of $a_{\text{ho}}/a_{3\text{D}}$.
(a) The symbols 
show the dimensionless energy difference 
$(E_3-E_{2}-\hbar \omega)/(\hbar \omega)$
for 
$\kappa=3/2$ (top curve; exists only 
around $a_{\text{ho}}/a_{3\text{D}} = -1$), $\kappa=2$ (middle curve) and 
$\kappa=5/2$
(bottom curve).
(b)
The symbols show the dimensionless energy difference 
$(E_3-E_{2}-\hbar \omega)/(\hbar \omega)$
for 
$\kappa=4$ (top curve), $5$, $13/2$, $8$, and $10$ (bottom curve).
The lines connect the data points as a guide to the eye.
The calculations are performed for $r_0=0.01a_{\text{ho}}$.
}
\label{fig1_symm2}
\end{figure}

Figure~\ref{fig1_symm2}
shows the three-body energies for 
the lowest state with
$(\Pi_{\vec{\rho}},M_{\text{rel}},\Pi_z,a_y)=(+1,0,-1,+1)$ symmetry as a function of
$a_{\text{ho}}/a_{3\text{D}}$ for
various mass ratios.
The results are obtained for $r_0=0.01a_{\text{ho}}$.
As mentioned above, three-body states are bound with respect 
to the breakup into
a dimer and an atom if their energy is less than 
$E_{2}+\hbar \omega$.
Correspondingly, Fig.~\ref{fig1_symm2} shows the quantity
$(E_3-E_{2}-\hbar \omega)/(\hbar \omega)$.

For $r_0=0.01a_{\text{ho}}$,
the system with $\kappa=3/2$ supports three-body bound states
with binding energies around $-6.5 \times 10^{-5} \hbar \omega$ and
$-2 \times 10^{-4} \hbar \omega$ (these are variational upper bounds)
for
$a_{\text{ho}}/a_{3\text{D}}=-1.5$ and $-1$, respectively.
For the range of $r_0=0.01a_{\text{ho}}$, we find no three-body bound states
for $a_{\text{ho}}/a_{3\text{D}}=-2$ and $-0.5$. 
While the exact threshold scattering lengths, i.e., the scattering lengths at
which the system becomes unbound, depend on the range of the underlying 
two-body potential, our results confirm the existence of
weakly-bound states for $\kappa>1$.

\begin{figure}
\vspace*{.5cm}
\includegraphics[angle=0,width=65mm]{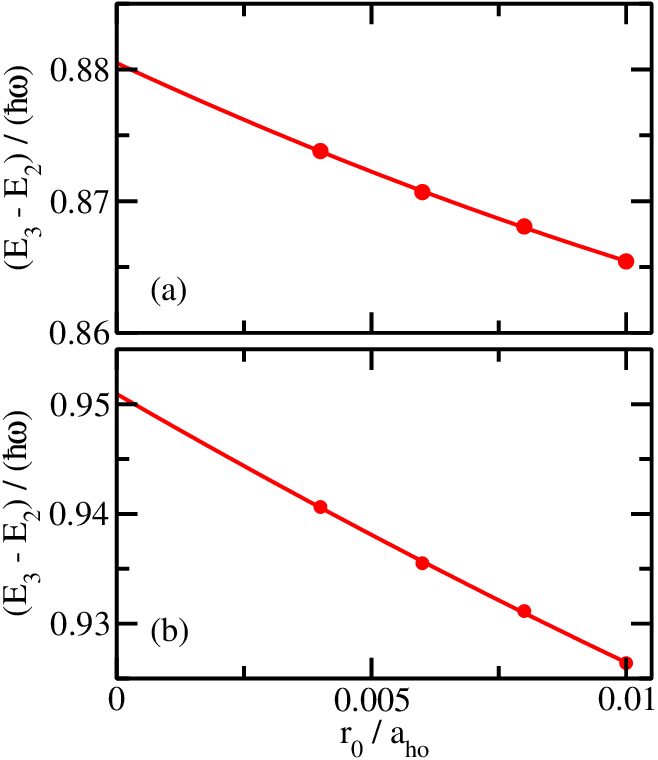}
\vspace*{.5cm}
\caption{(Color online)
Range dependence of the relative three-body energy of the lowest state
with
$(\Pi_{\vec{\rho}},M_{\text{rel}},\Pi_z,a_y)=(+1,0,-1,+1)$ symmetry
and $\kappa=5$.
The symbols show the 
dimensionless energy difference 
$(E_3-E_{2})/(\hbar \omega)$ 
for
(a) $a_{\text{ho}}/a_{3\text{D}}=1$ and
(b) $a_{\text{ho}}/a_{3\text{D}}=2$.
The lines show three parameter fits to the scaled finite-range energies.
}
\label{fig2_symm2}
\end{figure}

For larger mass ratios, the scattering length window for which
three-body bound states are supported increases, especially on
the negative scattering length side. On the positive scattering length side,
the bound state region also increases.
For $\kappa=5$, e.g., three-body bound states are supported 
for positive $g$ (i.e., for $a_{\text{ho}}/a_{3\text{D}} \gtrsim 1/1.03263$).
To see if this is a consequence 
of the finite-range nature of the interactions,
Figs.~\ref{fig2_symm2}(a) and \ref{fig2_symm2}(b)
show the scaled three-body energy 
$(E_3-E_{2})/(\hbar \omega)$
as a  
function of $r_0$ for $a_{\text{ho}}/a_{3\text{D}}=1$ and $a_{\text{ho}}/a_{3\text{D}}=2$, respectively. 
Although the binding energy decreases
with decreasing range, Fig.~\ref{fig2_symm2} shows 
that the three-body system is bound for all $r_0$
considered, including the zero-range limit.
This implies that three-body bound states are, for sufficiently large 
$\kappa$, not only supported if $g$ is negative but also if $g$ is positive.
This is in contrast to the prediction
based on the purely one-dimensional framework~\cite{kart09}, 
where a positive $g$ 
corresponds to a 
purely repulsive system.
The strictly one-dimensional treatment could be improved, as suggested
in Refs.~\cite{olshanii,moore}, by using the energy-dependent
Hurwitz zeta function instead of the 
energy-independent zeta function in the
parametrization of the one-dimensional coupling constant $g$
[see Eq.~(\ref{eq_g1d})].

\begin{figure}
\vspace*{+1.5cm}
\includegraphics[angle=0,width=65mm]{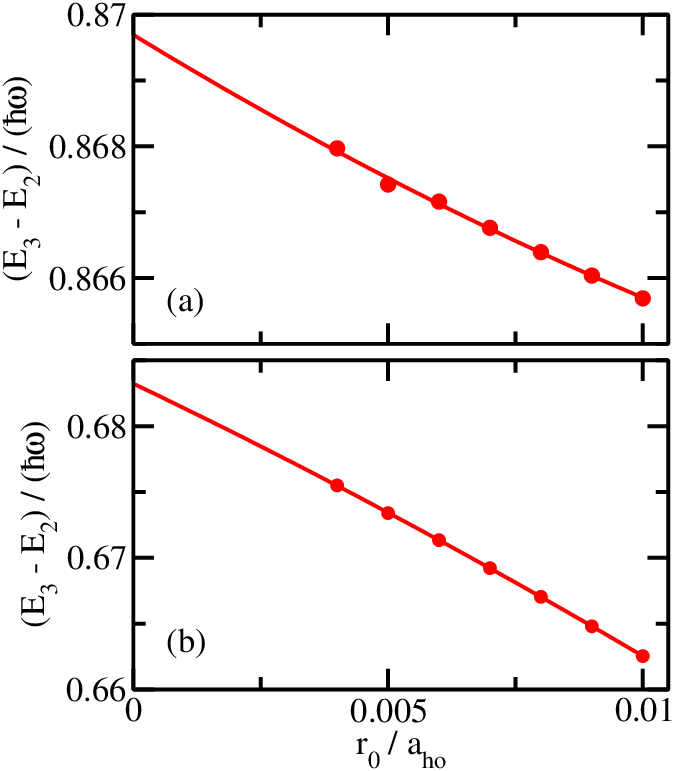}
\vspace*{0.5cm}
\caption{(Color online)
Range dependence of the relative three-body energy of the lowest state
with
$(\Pi_{\vec{\rho}},M_{\text{rel}},\Pi_z,a_y)=(+1,0,-1,+1)$ symmetry
and $a_{\text{ho}}/a_{3\text{D}}=0$.
The symbols show
the dimensionless
energy difference 
$(E_3-E_{2})/(\hbar \omega)$ 
for
(a) $\kappa=6$ and (b) $\kappa=10$.
The lines show three parameter fits to the scaled finite-range energies.
}
\label{fig3_symm2}
\end{figure}

\begin{figure}
\vspace*{+.5cm}
\includegraphics[angle=0,width=65mm]{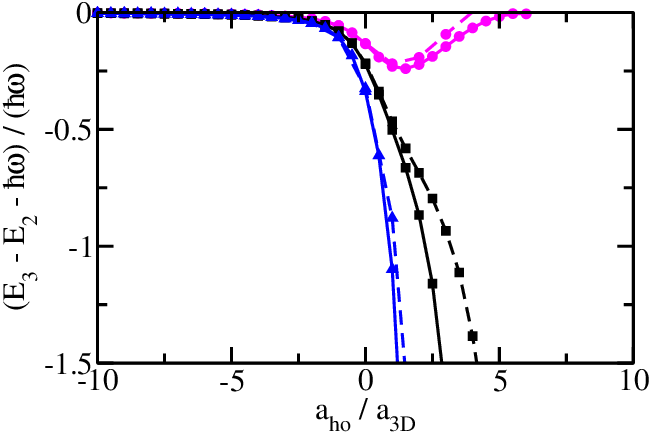}
\vspace*{0.5cm}
\caption{(Color online)
Range-dependence of the
relative three-body energies for the lowest state with
$(\Pi_{\vec{\rho}},M_{\text{rel}},\Pi_z,a_y)=(+1,0,-1,+1)$
symmetry as a function of $a_{\text{ho}}/a_{3\text{D}}$.
The circles, squares and triangles 
show the dimensionless energy difference 
$(E_3-E_{2}-\hbar \omega)/(\hbar \omega)$
for $\kappa=6, 8$ and $10$, respectively.
The dimensionless energies are connected by dashed and solid lines
for
$r_0=0.005a_{\text{ho}}$ and $0.01a_{\text{ho}}$, respectively.
}
\label{fig4_symm2}
\end{figure}

To determine whether the three-body bound states
are universal 
for larger $\kappa$ and large $|a_{3\text{D}}|$,
Figs.~\ref{fig3_symm2}(a) and \ref{fig3_symm2}(b) show
the range dependence of the scaled three-body energy
for $a_{\text{ho}}/a_{3\text{D}}=0$ as a function of
$r_0$ for $\kappa=6$ and $10$, respectively.
The range dependence is quite small even for these large mass ratios, 
suggesting that the three-body bound states
in the $(\Pi_{\vec{\rho}},M_{\text{rel}},\Pi_z,a_y)=(+1,0,-1,+1)$
channel are rather insensitive to the details of the underlying
two-body potential if 
$a_{\text{ho}}/a_{3\text{D}} \ll 1$,
$a_{\text{ho}}/r_0 \gg 1$ and 
$|a_{3\text{D}}|/r_0 \gg 1$.
In the regime where $a_{\text{ho}}/a_{3\text{D}} \gg 1$, however,
the finite-range effects are 
notably more important.
As an example, Fig.~\ref{fig4_symm2} compares the
scaled energy $(E_3-E_{2}-\hbar \omega)/(\hbar \omega)$
for $r_0=0.01a_{\text{ho}}$ (solid lines) and $r_0=0.005a_{\text{ho}}$ 
(dashed lines) for $\kappa=6, 8$ and $10$.
While the qualitative behavior is independent of $r_0$, quantitative differences
are visible in the $a_{\text{ho}} > a_{3\text{D}}$ regime.

\begin{figure}
\vspace*{+.5cm}
\includegraphics[angle=0,width=65mm]{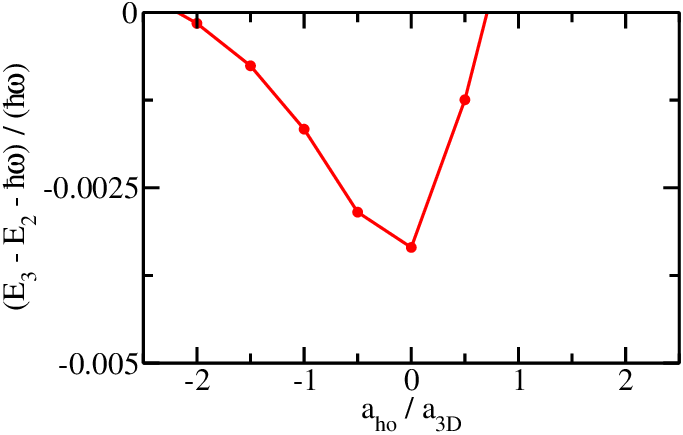}
\vspace*{0.5cm}
\caption{(Color online)
Relative energy of the first excited three-body state 
with
$(\Pi_{\vec{\rho}},M_{\text{rel}},\Pi_z,a_y)=(+1,0,-1,+1)$ symmetry as 
a function of $a_{\text{ho}}/a_{3\text{D}}$.
The symbols show the scaled energy
$(E_3-E_{2}-\hbar \omega)/(\hbar \omega)$ for
$\kappa=9$ and $r_0=0.01a_{\text{ho}}$.
Lines connect the data points as a guide to the eye.
}
\label{fig5_symm2}
\end{figure}

Lastly, we search for excited states in the 
$(\Pi_{\vec{\rho}},M_{\text{rel}},\Pi_z,a_y)=(+1,0,-1,+1)$ channel.
As discussed above, the strictly one-dimensional framework predicts that
excited states are supported 
if $\kappa$ is greater than $7.3791$~\cite{kart09}.
For $\kappa=8$ and $r_0=0.01a_{\text{ho}}$, we found that the 
first excited state is,
within our variational treatment, not bound with respect to the break-up into
a dimer and an atom.
We did not investigate how this ``negative result''
depends on the range of the underlying two-body potential.
For $\kappa=9$, however, the system supports an excited three-body bound state,
as expected from the strictly one-dimensional framework.
Circles in Fig.~\ref{fig5_symm2}
show the scaled energy of the first excited state as a function
of $a_{\text{ho}}/a_{3\text{D}}$ for $\kappa=9$ and $r_0=0.01a_{\text{ho}}$. 
The dependence of the excited states
on the $s$-wave scattering length seems to be 
similar to that of the ground state (compare 
Fig.~\ref{fig5_symm2} with Figs.~\ref{fig1_symm2} and
\ref{fig4_symm2}).

\subsection{$(\Pi_{\vec{\rho}},|M_{\text{rel}}|,\Pi_z)=(-1,1,+1)$}
\label{sec_symm3}
\begin{figure}
\vspace*{+.5cm}
\includegraphics[angle=0,width=65mm]{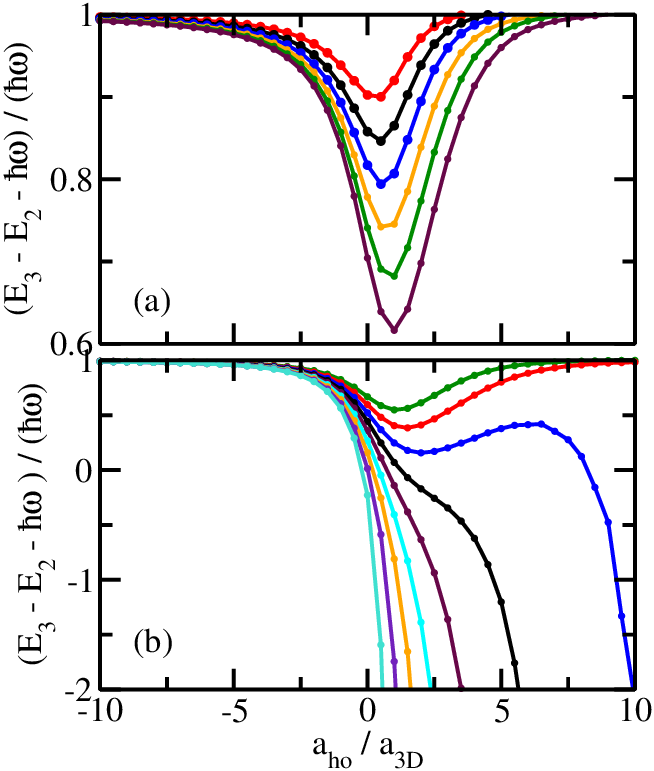}
\vspace*{0.5cm}
\caption{(Color online)
Relative three-body energies for 
the lowest state with $(\Pi_{\vec{\rho}},|M_{\text{rel}}|,\Pi_z)=(-1,1,+1)$
symmetry as a function of $a_{\text{ho}}/a_{3\text{D}}$.
(a) Symbols show the dimensionless energy difference 
$(E_3-E_{2}-\hbar \omega)/(\hbar \omega)$
for 
$\kappa=1$ (top curve), $3/2$, $2$, $5/2$, $3$ and $7/2$ (bottom curve).
(b)
The symbols show the dimensionless energy difference 
$(E_3-E_{2}-\hbar \omega)/(\hbar \omega)$
for 
$\kappa=4$ (top curve), $5$, $6$, $7$, $8$, $9$, $10$, $11$ and $12$
(bottom curve).
The lines connect the data points as a guide to the eye.
The calculations are performed for $a_0=0.01a_{\text{ho}}$.
}
\label{fig1_symm3}
\end{figure}

This section explores under which conditions
the three-body system in 
the $(\Pi_{\vec{\rho}},|M_{\text{rel}}|,\Pi_z)=(-1,1,+1)$
channel supports bound states that are stable with respect to the
lowest dimer plus atom threshold with energy $E_{2}+\hbar \omega$.
Figure~\ref{fig1_symm3} shows the dimensionless
energy $(E_3-E_{2}-\hbar \omega)/(\hbar \omega)$
for $r_0=0.01a_{\text{ho}}$
as a function of $a_{\text{ho}}/a_{3\text{D}}$ for various $\kappa$.
For $\kappa=1$ [top curve in Fig.~\ref{fig1_symm3}(a)],
the scaled energy 
$(E_3-E_{2}-\hbar \omega)/(\hbar \omega)$ 
shows a minimum near 
$a_{\text{ho}}/a_{3\text{D}}=0$. 
As $\kappa$ increases, the minimum deepens and 
moves slightly to the positive scattering length side.
While these systems with $\kappa$ not much larger than 1 
are bound with respect to
the excited dimer plus atom threshold with energy $E_{2}+2 \hbar \omega$,
they are not bound with respect to the lowest 
dimer plus atom threshold with energy $E_{2}+\hbar \omega$.
For $\kappa=6$ [third curve from the top in
Fig.~\ref{fig1_symm3}(b)], 
the scaled energy 
$(E_3-E_{2}-\hbar \omega)/(\hbar \omega)$ 
drops below $-1$ for large $a_{\text{ho}}/a_{3\text{D}}$.
For yet larger $\kappa$, the scaled energy
$(E_3-E_{2}-\hbar \omega)/(\hbar \omega)$ 
decreases monotonically with increasing $a_{\text{ho}}/a_{3\text{D}}$.
For $\kappa=8-12$, the three-body system 
with $r_0=0.01 a_{\text{ho}}$ becomes bound with 
respect to the lowest dimer plus atom
threshold for $a_{\text{ho}}/a_{3\text{D}}$ between approximately $3$ to $0.5$.

\begin{figure}
\vspace*{+1.5cm}
\includegraphics[angle=0,width=65mm]{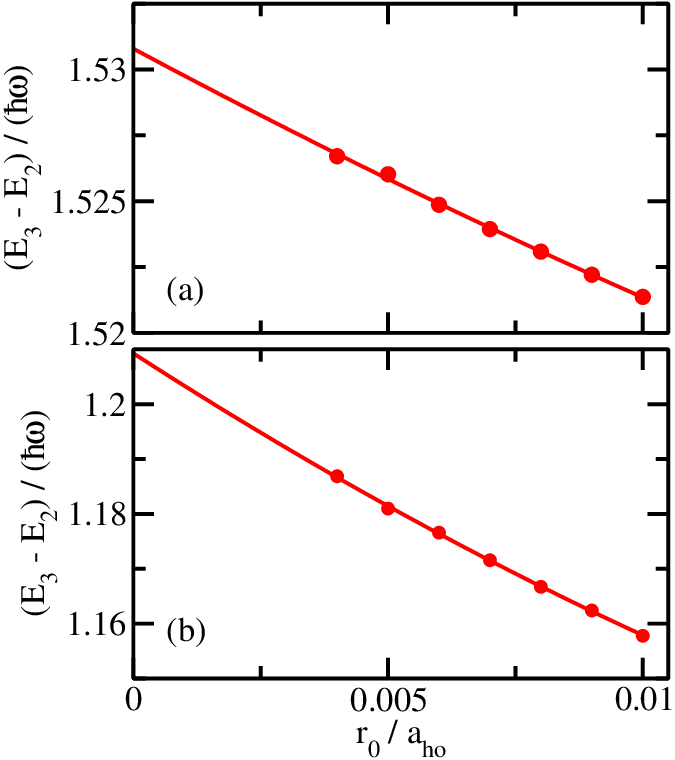}
\vspace*{0.5cm}
\caption{(Color online)
Range dependence of the relative three-body energy of the lowest state
with
$(\Pi_{\vec{\rho}},|M_{\text{rel}}|,\Pi_z)=(-1,1,+1)$ symmetry
and  $a_{\text{ho}}/a_{3\text{D}}=0$.
The symbols show
the dimensionless
energy difference $(E_3-E_{2})/(\hbar \omega)$ for
(a) $\kappa=6$ and (b) $\kappa=10$.
The solid lines show three parameter fits to the scaled finite-range energies.
}
\label{fig2_symm3}
\end{figure}

To investigate the range dependence of the three-body energies,
Figs.~\ref{fig2_symm3}(a) and 
\ref{fig2_symm3}(b) show the energy difference
$(E_3-E_{2})/(\hbar \omega)$
for $a_{\text{ho}}/a_{3\text{D}}=0$
as a function of $r_0$
for $\kappa=6$ and $\kappa=10$, respectively.
The energy difference depends
approximately linearly on the range.
Figures~\ref{fig2_symm3}(a) and \ref{fig2_symm3}(b)
show that the
range dependence increases with increasing $\kappa$.

\begin{figure}
\vspace*{+.5cm}
\includegraphics[angle=0,width=65mm]{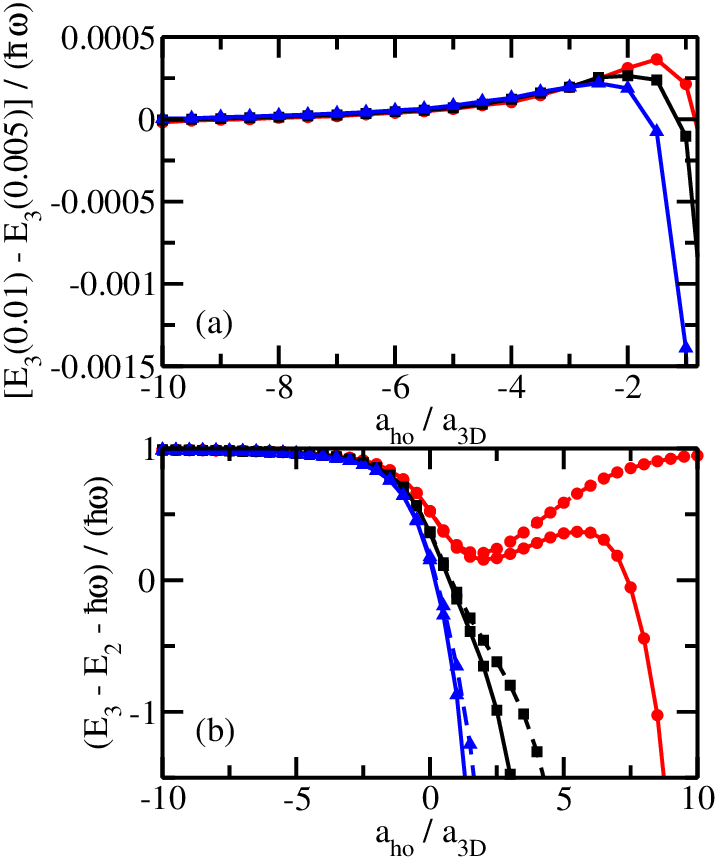}
\vspace*{0.5cm}
\caption{(Color online)
Range dependence of the relative three-body energy of the lowest state
with
$(\Pi_{\vec{\rho}},|M_{\text{rel}}|,\Pi_z)=(-1,1,+1)$ symmetry
as a function
of $a_{\text{ho}}/a_{3\text{D}}$.
(a) The circles, squares and triangles show the
energy difference 
$[E_3(r_0=0.01a_{\text{ho}})-E_3(r_0=0.005a_{\text{ho}})]/( \hbar \omega)$
for $\kappa=6$, 8 and 10, respectively.
The lines connect the data points as a guide to the eye.
(b) The circles, squares and triangles
show the quantity $(E_3-E_{2}-\hbar \omega)/(\hbar \omega)$
for $\kappa=6$, 8 and 10.
The dimensionless energies are connected by
dashed and solid lines for
$r_0=0.005a_{\text{ho}}$ and
$r_0=0.01a_{\text{ho}}$, respectively.
}
\label{fig3_symm3}
\end{figure}

To obtain a sense of the range dependence on the negative scattering length
side,
Fig.~\ref{fig3_symm3}(a) shows the difference 
between the three-body energies for $r_0=0.01a_{\text{ho}}$ and 
$r_0=0.005a_{\text{ho}}$ for $\kappa=6$ (circles),
$\kappa=8$ (squares) and $\kappa=10$ (triangles).
The range dependence is very small on the negative scattering length 
side. 
To visualize the range dependence in the strongly interacting regime
(including the positive scattering length side),
Fig.~\ref{fig3_symm3}(b) shows the scaled energies 
$(E_3-E_{2}-\hbar \omega)/(\hbar \omega)$
for two different ranges, $r_0=0.01a_{\text{ho}}$ and $r_0=0.005a_{\text{ho}}$,
and three mass ratios, $\kappa=6$, 8 and 10.
Roughly speaking, the range of the two-body potential
becomes important when $a_{\text{ho}}/a_{3\text{D}} \gtrsim -1$.

As pointed out earlier, the states with 
$(\Pi_{\vec{\rho}},|M_{\text{rel}}|,\Pi_z)=(-1,1,+1)$ symmetry
considered here are consistent with the $(L,\Pi)=(1,-1)$
symmetry of the three-dimensional free-space system.
For mass ratios $\kappa> 8.173$ and zero-range
interactions,
the energy of the free-space 
system in the $(1,-1)$ channel is directly proportional
to $(a_{3{\text{D}}})^{-2}$.
Thus we expect that the three-body energies for the wave guide Hamiltonian 
for positive $a_{3{\text{D}}}$ scale in the same way.
We find that this is only approximately true for the parameter
regime explored in this work.
The requirement that $a_{3\text{D}}$ should be less than $a_{{\text{ho}}}$
and larger than $r_0$, combined with large finite range effects, make 
it challenging,
at least for the numerical approach employed in this work,
to reach the regime where the energies for the wave guide
Hamiltonian approach those for the 
free-space Hamiltonian with zero-range interactions.

\section{Summary}
\label{sec_summary}

This paper determined the bound states of two
identical heavy fermions and one light particle in a harmonic waveguide
for short-range interspecies $s$-wave interactions.
Our calculations accounted for the full dynamics along the 
direction of the harmonic confinement
as well as along the direction of the waveguide, i.e., coupling between the
degrees of freedom along these directions was treated explicitly.
Comparisons with predictions based on an effective one-dimensional 
Hamiltonian were presented.
We investigated three different symmetries:

{\em{(i)}}
For states with $(\Pi_{\vec{\rho}},|M_{\text{rel}}|,\Pi_z,a_y)=(+1,0,+1,+1)$
symmetry,
no three-body bound states were found for the mass ratios investigated.
This finding is in agreement with what is expected based on results for 
an effective one-dimensional Hamiltonian and the three-dimensional 
free-space results for $(L,\Pi)=(0,+1)$
symmetry.

{\em{(ii)}}
For states with $(\Pi_{\vec{\rho}},|M_{\text{rel}}|,\Pi_z,a_y)=(+1,0,-1,+1)$
symmetry,
three-body bound states were found for $\kappa>1$ in the strongly-interacting
regime.
For sufficiently large $\kappa$, three-body bound states exist not only 
on the negative scattering length side but also on the
positive scattering length side.
While the bound states on the
positive scattering length side are absent in the strictly one-dimensional
treatment, their existence for sufficiently large $\kappa$
is expected since 
free-space systems with $(L,\Pi)=(1,-1)$ symmetry
support universal three-body states for positive $a_{3\text{D}}$ and
$\kappa>8.173$~\cite{kart07}.

{\em{(iii)}}
For states with $(\Pi_{\vec{\rho}},|M_{\text{rel}}|,\Pi_z)=(-1,1,+1)$
symmetry,
three-body bound states were found for sufficiently large $\kappa$.
This is a new class of bound states that has, to the best of our knowledge,
not been considered before.
The anti-symmetry of the corresponding  eigenstates is ensured
by placing an excitation into the angular degrees of freedom,
allowing the solution
along the waveguide axis to have positive parity (i.e., $\Pi_z=+1$)
and no nodes.
The three-body 
bound state first emerges on the positive scattering length side.

A variety of unequal-mass systems have been trapped and cooled to
the degenerate or near-degenerate regime over the past 10 years or so,
and the creation of effectively one-dimensional confining geometries 
is fairly standard by now.
Recent experiments on K-Li mixtures with mass ratio
$\kappa \approx 6.5$~\cite{coldatoms5}, e.g., investigated the
effects of the $L=1$ states on the positive $s$-wave scattering
length side on the collision dynamics
in the three-dimensional regime.
It would be very interesting to extend these experimental
studies to the effectively one-dimensional
regime, where the strength of the confinement can
be used to tune the interaction strength.
By changing $\omega$, the ratio $a_{\text{ho}}/a_{3\text{D}}$ and,
correspondingly, the 
position of the three-body bound state relative to the
lowest dimer plus atom threshold can be tuned. It would be interesting to
monitor the three-body recombination rate and to thus
indirectly search for signatures of the three-body bound states in the 
$(\Pi_{\vec{\rho}},|M_{\text{rel}}|,\Pi_z,a_y)=(+1,0,-1,+1)$
and 
$(\Pi_{\vec{\rho}},|M_{\text{rel}}|,\Pi_z)=(-1,1,+1)$
channels.
 Alternatively, 
it would be interesting to probe the three-body bound states
directly by radio-frequency spectroscopy.
In the future, it will be interesting to extend the studies presented here 
to
other confinement geometries, to other particle symmetries and to
larger systems.

{\em{Acknowledgement:}}
DB is grateful to Janine Shertzer for extensive discussions
involving the symmetry of the Hamiltonian and for preliminary 
calculations of effective hyperradial potential curves
using a 4D finite element analysis.
DB also thanks Debraj Rakshit and
Ebrahim Gharashi for helpful discussions,
and acknowledges
support by the NSF through
grant PHY-1205443.
This work was additionally supported by the National Science
Foundation through a grant for the Institute for
Theoretical Atomic, Molecular and Optical Physics
at Harvard University and Smithsonian Astrophysical Observatory.


\begin{thebibliography}{100}
\bibitem{footnotereflection}
Instead of defining the reflection with respect to the $y$-axis, we
could reflect with respect to the $x$-axis or any axis
that lies in the $xy$-plane and
goes through the origin.
\end{thebibliography}
\end{document}